# Resolving Single-Peptide Phosphorylation Dynamics in Plasmonic Nanopores using Physics-Informed Bi-Path Model

Mulusew W. Yaltaye[1,2,3], Yingqi Zhao[1], Kuo Zhan[1], Vahid Farrahi[1,4], Jian-An Huang [1,2,3,]*

[1] Research Unit of Health Sciences and Technology, Faculty of Medicine, University of Oulu, Aapistie 5 A, 90220 Oulu, Finland.

[2] Research Unit of Disease Networks, Faculty of Biochemistry and Molecular Medicine, University of Oulu, Aapistie 5 A, 90220 Oulu, Finland.

[3] Biocenter Oulu, University of Oulu, Aapistie 5 A, 90220 Oulu, Finland.

[4] Institute for Sports and Sport Sciences, TU Dortmund University, Germany

*Email: jianan.huang@oulu.fi

**Abstract**

Protein phosphorylation provides a dynamic readout of cellular signaling yet remains difficult to detect at low abundance and stoichiometry. Single-molecule surface-enhanced Raman spectroscopy (SM-SERS) using particle-in-pore plasmonic nanopores offers label-free molecular detection with submolecular sensitivity. However, reliable identification of subtle post-translational modifications (PTMs) is hindered by the stochastic nature of SM-SERS signals, partial excitation of peptide residues within the plasmonic hotspot, and background interference. Here, we introduce a physics-informed deep learning framework to decode complex SM-SERS dynamics and identify single-peptide PTMs. The model integrates multiple-instance learning with a temporal encoder combining temporal convolutional networks and bidirectional gated recurrent units to capture both local spectral variability and long-range blinking dynamics. To address diffusion-driven spectral heterogeneity, long spectral trajectories are segmented using Pearson-correlation, enabling weakly supervised training under label ambiguity. This framework robustly distinguishes single peptide phosphorylation despite strong background interference and stochastic signal fluctuations. By coupling nanoplasmonic confinement with spatiotemporal deep learning, our approach enables high-fidelity detection of single-molecule phosphorylation events and advances ultrasensitive phosphoproteomic analysis.

**Introduction**

Protein phosphorylation provides a functional snapshot of signaling dynamics, yet it remains difficult to quantify at low abundance and low stoichiometry [1–3]. Consequently, scientist continues to push towards ultrasensitive phosphoproteomics and developing new single-molecule approaches. While mass spectrometry remains the gold standard for proteomic analysis, its utility is constrained by a requirement of millions copies of molecules [1,4]. The emerging landscape of single-molecule protein sequencing technologies offers a promising alternative to overcome these sensitivity limits. However, these technologies still face significant hurdles, including low throughput and complex sample preparation [5]. Fluorescence-based methods are another alternatives for single molecule sequencing, but they require labeling with fluorescent tags and suffer from limited labelling and photobleaching, which in turn limits the read length [5,6]. Recent studies have demonstrated that engineered nanopores can achieve accurate molecular detection of peptides through rational design and optimization of pore structural properties [7,8]. However, current nanopore approaches primarily rely on ionic current measurements, leaving room for innovative sensing modalities that could provide enhanced specificity and spatial information.

Single-molecule surface-enhanced Raman spectroscopy (SM-SERS) has emerged as a revolutionary analytical technique capable of detecting individual molecules in a non-invasive, label-free manner with high throughput [9]. SM-SERS transcends traditional ensemble measurements to reveal molecular heterogeneity and dynamics previously obscured by statistical averaging of SERS spectra of multi-molecules [10,11]. The Particle-in-pore Sensor represents one of the most powerful SM-SERS detection platforms, capable of detecting single-molecule spectrum by trapping a gold nanoparticle (AuNP) inside a nanopore as shown in **Fig. 1A** [12,13]. When illuminated at its resonance wavelength, the optical near field is confined into highly intense localized region so-called "hotspot." The molecule diffuses through, or transiently bind within, this hotspot experiencing orders-of-magnitude enhancement of their Raman scattering, enabling single molecule detection (**Fig. 1B**). The particle-in-pore plasmonic sensor is particularly attractive for biomolecular sensing, as it allows label-free readout of molecular structure in solutions while preserving native conformational dynamics [14,15].

Despite these advantages, the inherently stochastic nature [16,17] of SM-SERS signals poses a significant analytical challenge, as single-molecule spectra exhibit pronounced temporal fluctuations and intermittent "blinking" driven by Brownian motion of the analyte on the surface of AuNPs and dynamic reconfiguration of the plasmonic field (**Fig. 1C**) [18,19]. Spectral fluctuations are further complicated by background interference, particularly from citrate commonly used as a stabilizing agent to the gold nanoparticles, which can compete with the analyte for occupancy in the hotspot and obscure low-abundance spectral features [12,13]. In the case of single peptide PTM analysis, the analysis becomes significantly more complex due to partial excitation and overlapped sequences. The unphosphorylated peptide segment F-Ser (CDSSPDSAEDVRK) and the phosphorylated peptide F-pSer (CDSSPDpSAEDVRK), both derived from Fetuin-A (FETUA/AHSG) [20,21], share many amino acids, while a hotspot may interrogate only 1–3 amin acid residues at a time. In addition to this, aromatic amino acid residue dominates over non-aromatic as Raman cross-sections vary substantially[4]. These combined constraints render phosphorylation identification at a peptide level very challenging, because the phosphate group often results in subtle conformational changes or minor spectral shifts that are easily eclipsed by the surrounding peptide backbone[22].

To address these physical constraints in SM-SERS, we herein developed a physics-informed, bi-path hierarchical model to decode single-peptide PTM dynamics (**Fig. 1D**). The model integrates Multiple Instance Learning (MIL) with advanced temporal encoding. Recently, several supervised machine learning and deep learning techniques have been utilized to analyze Raman spectroscopic data[13,23–25]. However, traditional supervised learning algorithms, which typically require pixel-level annotations, struggles when applied to SM-SERS peptide data acquired in flow-through or dynamic setups[26,27]. This limitation arises because labels are assigned to entire time series, whereas individual spectra ("instances") fluctuate significantly due to Brownian motion, molecular rotation, and transient hotspot occupancy. Consequently, a single time series measurement contains a complex mixture of neighboring amino acid residues, idle states, and citrate-interference, which can be effectively addressed using Multiple Instance Learning (MIL). MIL is a weakly supervised learning framework that overcomes these limitations by naturally accommodating problems characterized by label ambiguity[28,29]. To prepare our data for MIL approach, we segment long time-series spectra into short segments using Pearson correlation, the approach considers the physical properties of peptide diffusion on the surface of AuNP in a plasmonic nanopore. Simultaneously, to capture the temporal evolution critical for distinguishing the subtle 'blinking' dynamics of F-Ser versus F-pSer, we employ a parallel Temporal Encoder comprising Temporal Convolutional Networks (TCNs) and bidirectional Gated Recurrent Units (BiGRUs), which captures both short-term molecular fluctuations and long-range dependencies without vanishing gradient issues[30,31]. By fusing these instance-level attention weights with sequence-level temporal dynamics in a hierarchical two-stage classifier, our model effectively learns molecular representations of transit through the plasmonic hotspot, enabling high-fidelity identification of post-translational modifications (PTMs) even within noisy and stochastic setups.

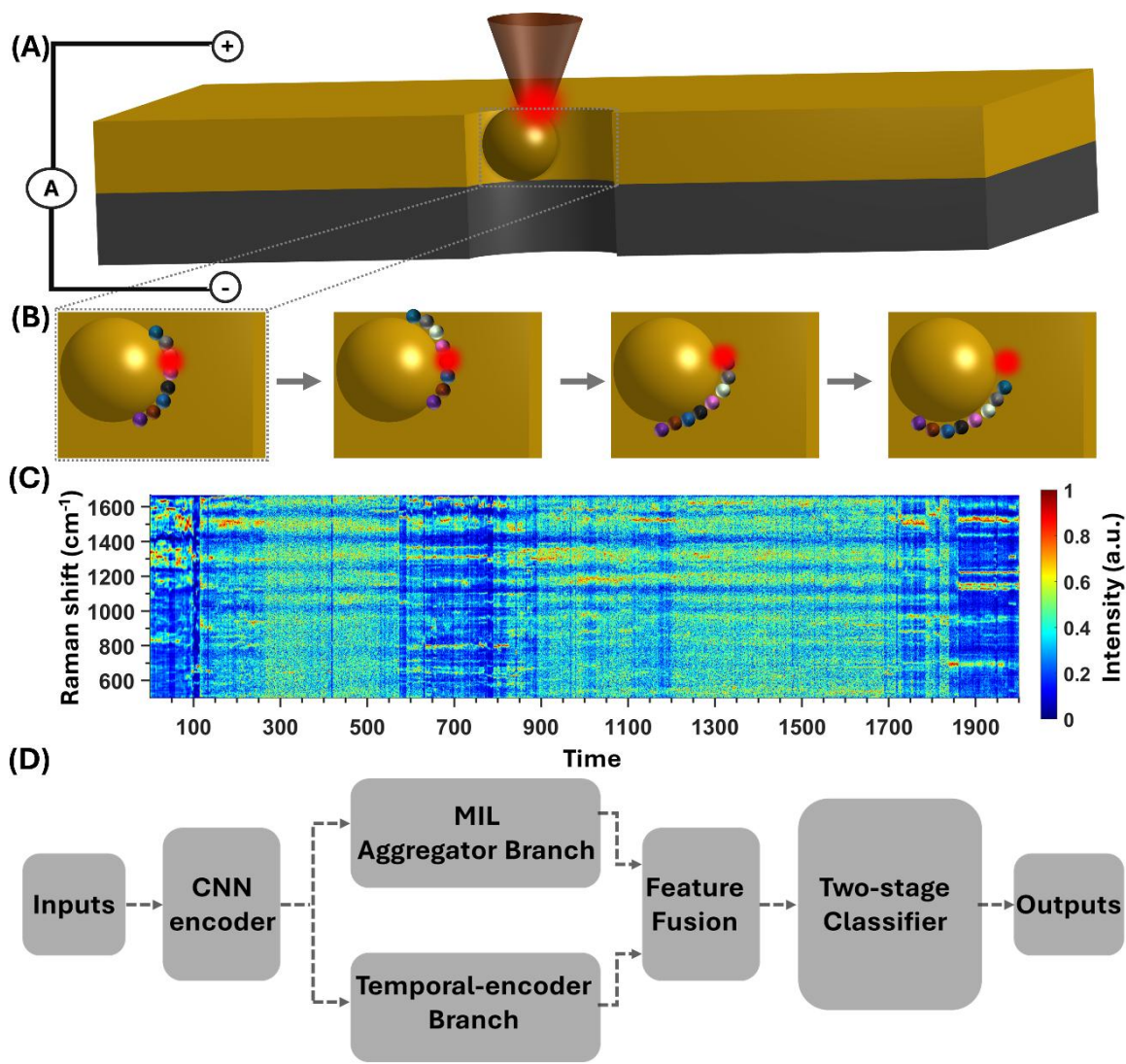

**Figure 1.** (A) Schematic of the plasmonic particle-in-pore sensor with a hot spot that excites part of single peptide. (B) the diffusion dynamics of molecules on the surface of gold-nanoparticles and the occupancy of the molecules the plasmonic hot spot. (C) the spectral map of single-peptide PTM to demonstrate variation in time, the color bar indicating the normalized intensity. (D) The overview of physics-informed bi-path model for disentangling single peptide PTM.

**Results and Discussion**

**Raman spectroscopic measurement of Single-molecule PTMs**

Plasmonic particle-in-pore sensor with single-molecule SERS sensitivity used to measure the single-molecule SERS spectra of PTMs according to the protocol in previous papers [4,19]. The gold nanopores of 200 nm diameter were fabricated on the silicon nitride ($Si_3N_4$) membrane by Focused ion Beam milling, then after the chip is encapsulated in a PDMS chamber to form cis/trans reservoirs. Target analytes physically adsorbed onto citrate-stabilized Au nanoparticles (AuNP) of 50 nm diameter to form a submonolayer (1/80 monolayer, approximately 1.23% surface coverage)[13] (see molecule concentrations in Supporting Information), the analyte-attached particle entrapped in the nanopore by

Page **4** of **23**

electro-opto-kinetic mechanism under 785 nm illumination the plasmonic nanopore generates a strong optical gradient force that draws the particle toward the sidewall, while an applied transmembrane bias balances electrophoretic and electroosmotic forces. ThermoFisher DXR2xi Raman microscope with 15 mW laser power, slit width of 50 μm, and 100 ms exposure time used to collect SERS spectra. Consequentely, a single plasmonic hot spots enabled the acquistion of SERS signals (**Fig. 1B**).

**Citrate-interference and shared spectral Regions**

The presence of citrate as stablizing agent of AuNPs introduces an interference in SM-SERS spectra of particle-in-pore-sensors that complicates data interpretation. In previous reports, citrate interference could be minimized by replacing citrate with a monolayer of analyte molecules through incubation for 48 hours[12]. In another approach, citrate interference was removed using k-means–based clustering by considering pure-citrate spectra as a reference and excluding any spectra that could contain citrate peaks[13]. The first method is limited by sensitivity, whereas the latter risks removing genuine signals from the target molecules, since the analyte may intrinsically share some spectral bands with citrate. The likelihood of overlapping spectral regions is even higher for peptides, since F-Ser (CDSSPDSAEDVRK) and F-pSer (CDSSPDpSAEDVRK) share the same peptide backbone, differing only by phosphorylation of the serine residue at the seventh position in the sequence (left to right). This results in shared spectral bands between molecules. As shown in **Figure 2**, a representative measurement demonstrates that all three molecules exhibit overlapping spectral bands around 715 $cm^{-1}$; F-pSer and citrate overlap at 1080 $cm^{-1}$; F-Ser and citrate at 1175 $cm^{-1}$; and F-Ser and F-pSer around 460 $cm^{-1}$, thereby further complicating the analysis.

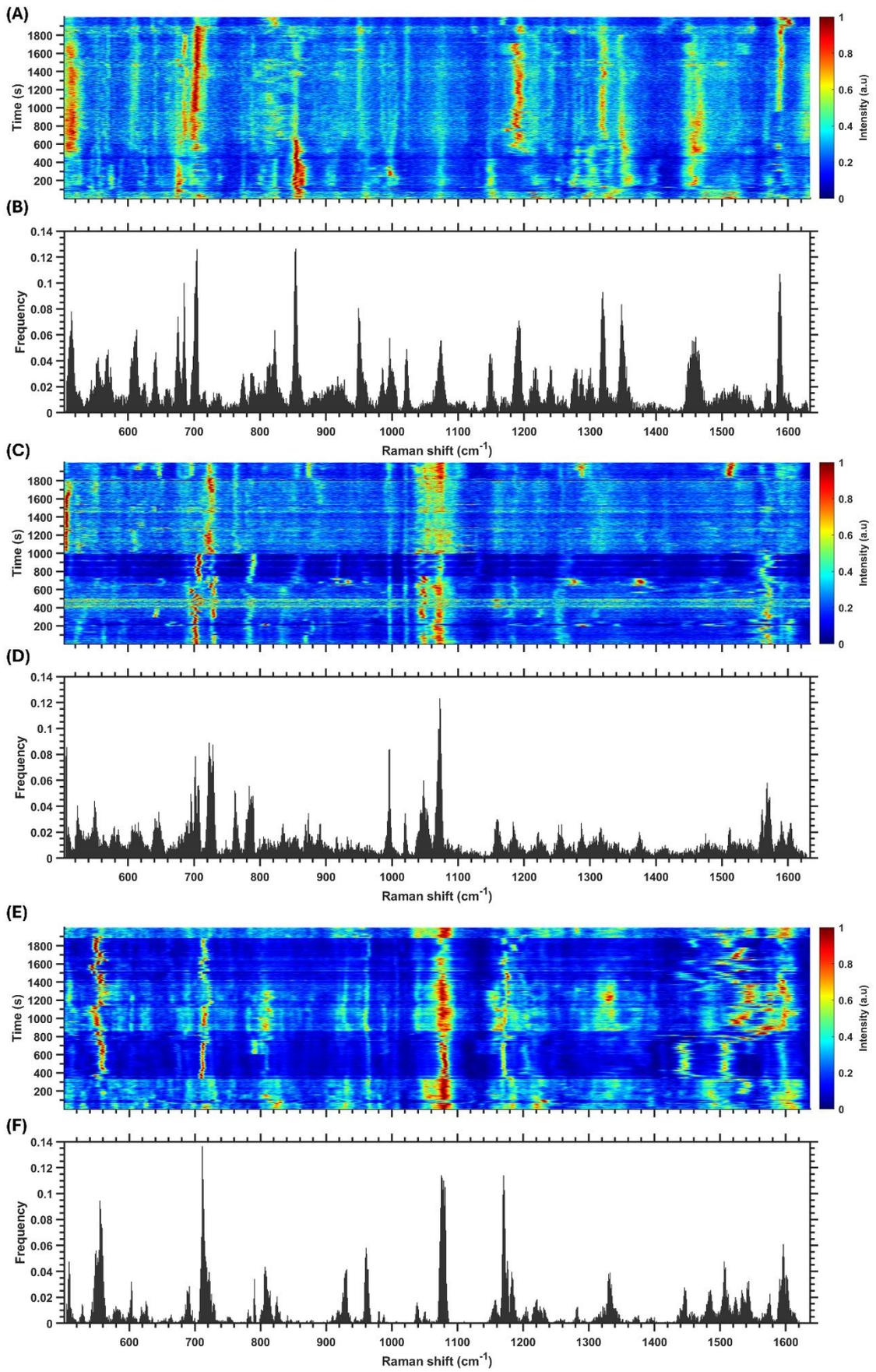

**Figure 2.** Spectral maps and corresponding peak-occurrence frequency distributions for F-Ser (a,b), F-pSer (c,d), and citrate (e,f). The spectral maps demonstrate overlapping spectral regions among the analyzed molecules, highlighting the challenge of distinguishing F-Ser from F-pSer.

Fluctuating peak, is one of the inherent properties of SM-SERS as shown in **Figure 2**, makes the traditional peak-assignment for individual spectra impractical. Thus, peak-occurrence frequency has become a valuable statistical tool for interpreting fluctuating single molecule spectra, where blinking and wandering often render individual spectra unrepresentative of the molecular signature. Peak-occurrence frequency summarizes a spectral time series by counting how frequently a vibrational band appears across many spectra and visualizing it as a normalized probability distribution over Raman shift, in which helps to emphasizes the most reproducible features/Raman shits of the molecule and effectively reconstructs a frequently occurring molecular fingerprint from transient and partial spectral snapshots arising from different trapping configurations. However, despite stabilizing highly variable SM-SERS signals, peak-occurrence analysis remains insufficient for distinguishing molecules with strongly overlapping spectral features.

**Locallized hotspot and Partial-Excitaion**

Notably, the hotspot of the particle-in-pore sensor **(Figure 1a)** excites only up to ~3 amino acid residues of a peptide at a time, depending on residue size, creating a fundamental challenge for identifying PTMs in long peptide or protein sequences[4,32]. Because the localized field illuminate only a fraction of the peptide, the resulting SERS spectra represent partial excitation of the peptide backbone rather than a complete vibrational fingerprint. Consequently, phosphate-associated vibrational modes may appear attenuated or intermittently absent, complicating residue-level assignment. In addition, random molecular motions, such as rotation and diffusion on the AuNP surface, can further obscure PTM sites. Addressing this partial-excitation limitation is therefore critical for advancing nanopore-enabled SERS toward robust, sequence-specific phosphoproteomics.

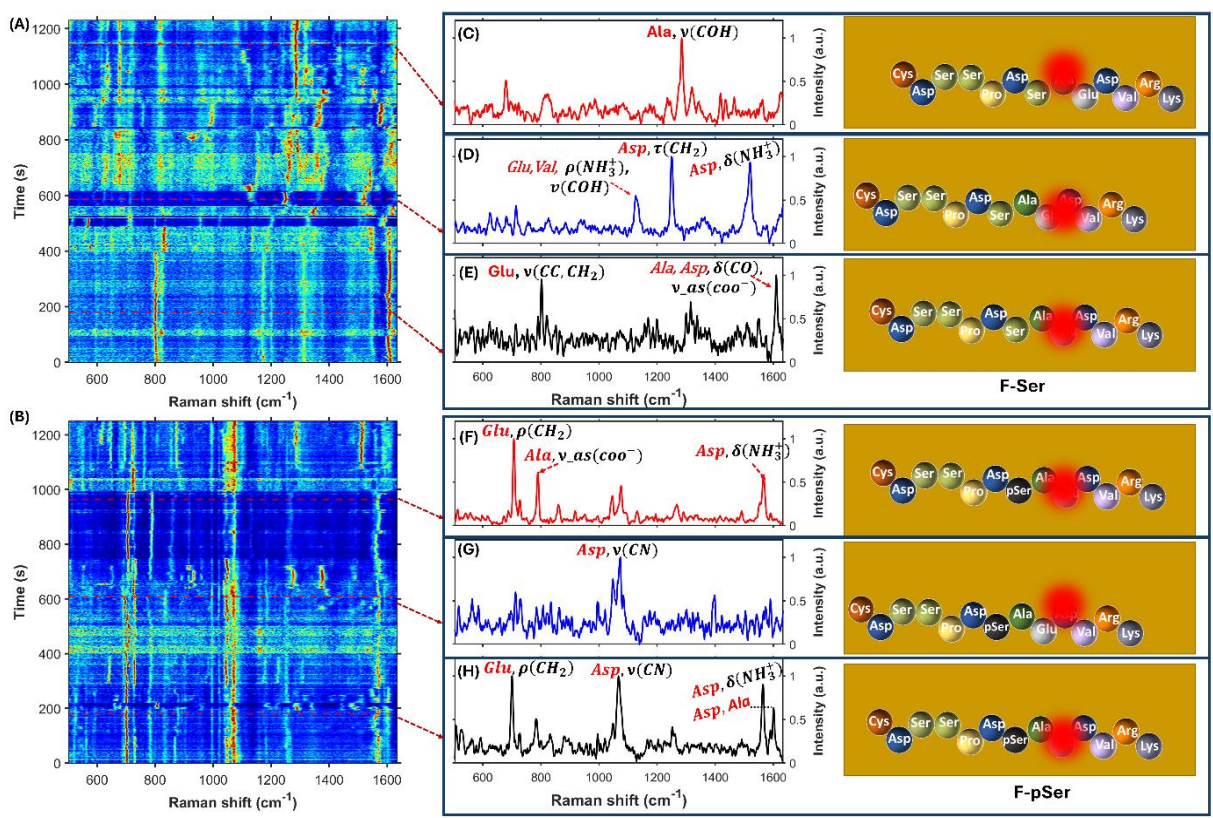

**Figure 3.** SERS time-series spectra of F-Ser (A) and F-pSer (B). Panels (C–E) show sample spectra of F-Ser at three different time points, together with assigned amino acid bases and their vibrational modes and the temporal hotspot configuration on peptide segment. Panels (F–H) show sample spectra of F-pSer at three different time points, along with their peak assignments and the corresponding hotspot-occupying amino acid residues based on the assignments.

Considering three spectra at different time from F-Ser, we have observed a vibrational band at 1280 $cm^{-1}$ is attributed to the C–O–H stretching mode (ν(COH)) from alanine in **Figure 3A**. In contrast, several vibrational modes are excited within the hotspot region (**Figure 3D**), including rocking and stretching of the ammonium group (ρ($NH_3^+$), ν(COH)), torsional motion (τ($CH_2$)), and bending modes (δ($NH_3^+$)) associated with glutamic acid, aspartic acid, and valine residues. In **Figure 3E**, the spectrum exhibits stretching and bending vibrational modes characteristic of glutamic acid, aspartic acid, and alanine. When these spectra are considered sequentially in time, they indicate that the peptide diffuses back and forth through the plasmonic hotspot, resulting in transient excitation of different residues. Similarly, for F-pSer, vibrational modes corresponding to glutamic acid, aspartic acid, and alanine are detected in **Figure 3F**, the signal is dominated by aspartic acid in **Figure 3G**; and in **Figure 3H**, glutamic acid, aspartic acid, and alanine again contribute to the detected vibrational modes. These observations suggest that the plasmonic hotspot excites only **1-3** amino acid bases. Detailed assignments of the corresponding vibrational modes are summarized in **Table 1**.

Tabel 1. Peak assignments of spectral bands from partial excitation of peptide backbones (Figure 3).

| Spectral band | Amino acid residues | Vibrational mode | References |
|---|---|---|---|
| 700 | Glu | $\rho(CH_2)$ | 33 |
| 720 | Glu | $\rho(CH_2)$ | 33 |
| 730 | Asp | $\rho(CH_2)$ | 34 |
| 795 | Ala | $v\_as(COO^-)$ | 35 |
| 800 | Glu | $v(CC), v(CH_2)$ | 33,36 |
| 1075 | Asp | $v(CN)$ | 34,37 |
| 1125 | Glu, Val | $\rho(NH_3^+), v(COH)$ | 33,38 |
| 1250 | Asp | $\tau(CH_2), v(COH)$ | 34,35,37 |
| 1280 | Ala | $v(COH)$ | 35 |
| 1520 | Asp | $\delta(NH_3^+)$ | 34,37 |
| 1575 | Asp | $\delta(NH_3^+)$ | 34 |
| 1600 | Ala, Asp | $\delta(CO), v\_as(COO^-)$ | 35,37,39 |

**Adaptive Segmentation using Pearson-correlation**

Peptide diffusion on the surface of AuNPs and the random entrapment of AuNPs within the nanopore result in short residence times of peptide segments in the hotspot, leading to stochastic spectral fluctuations, blinking, and partial excitation. Consequently, raw measurement labels represent a longer entrapment and broad classes rather than specific molecular states. To address this, we implemented a segmentation algorithm based on Pearson correlation, which used as a similarity metric to compare different spectra and identify the unique spectra from spectal datasets[40,41]. The algorithm scans time-series traces to identify signal onset, establishing a reference spectrum ($S_{ref}$) based on predefined criteria (see Supporting Information). Subsequent spectra are grouped into 'bags' (segments), if their correlation $r \geq 0.65$ with respect to ($S_{ref}$); otherwise, the current segment is terminated, and the divergent spectrum serves as the new ($S_{ref}$) for the subsequent iteration. This iterative approach mimics the physics of single molecules within a plasmonic hot spot. Physically, we assume a molecular resident-time of approximately 2.5 seconds (corresponding to 25 spectral records), after which molecular diffusion or nanoparticle displacement necessitates the detection of a new spectral signature to confirm the continued presence of a molecule (detailed Figure S2 in the supporting information). Pearson correlation is sensitive to the shape of the spectrum rather than absolute intensity thus it preclude the noisey and idle states[23,40,42].

**Deep Learning Model**

To accurately decode the complex single-peptide SERS data, we developed a Bi-Path hierarchical model. The model has two branches the spectral and temporal encoders. The pipeline begins with a shared 1D Convolutional Neural Network (CNN) that processes adaptively segmented spectral sequences. To emphasize subtle spectral variations e.g., peak shifts induced by phosphorylation, the input is engineered as a two-channel tensor comprising both the raw intensity and its first derivative, which has complementary role in subtle features extraction[43–45]. Applying a first-order derivative transformation captures spectral rate-of-change information helps remove baseline drift and other background interference while reducing overlap between spectral bands[46]. The CNN features are then fed into a channel-wise wavenumber attention mechanism, enabling the network to actively highlight discriminative Raman shifts while suppressing baseline noise. The resulting latent spectral embeddings are directed into two parallel processing pathways: a weakly supervised spatial aggregator and a temporal encoder.

The first pathway employs Multiple Instance Learning (MIL), a powerful weakly supervised paradigm specifically designed for data characterized by severe label ambiguity [47,48]. In single-molecule peptide data, a single measurement contains a complex mixture of transient molecular binding events, idle states, citrate interference, and partial hotspot excitations. Forcing frame-level annotations onto such stochastic data is impractical. Likewise, its success in isolating sparse diagnostic signals in clinical pathology [49–51], MIL bypasses the need for exhaustive pixel-level annotation. It utilizes gated attention to autonomously isolate and aggregate only the most informative, characteristics spectral "instances" into a robust, bag-level representation.

While the MIL branch is inherently order-invariant [52], the temporal evolution of the SERS signal contains vital physical information. To capture these dynamics, the second pathway utilizes a Temporal Convolutional Network (TCN) followed by a Bidirectional Gated Recurrent Unit (BiGRU). TCN employs causal dilated convolutions, preserving temporal ordering while expanding the receptive field to capture both short-lived spectral bursts and longer-range drift or binding events[53,54]. The BiGRU further integrates bidirectional dependencies, encoding forward and backward contextual information across the segment [55,56]. Finally, a temporal attention pooling layer compresses the sequence into a single feature vector that highlights the most meaningful kinetic transitions.

The complementary spatial (MIL) and kinetic (Temporal) representations are fused and passed to a two-stage hierarchical classifier. Mirroring the biochemical decision-making process, this classifier first separates the control background (Citrate) from target signals and subsequently distinguishes between the specific peptide variants (F-Ser vs. F-pSer). This hierarchical approach prevents dominant background variability from obscuring the subtle spectral differences of phosphorylation.

The entire architecture is trained end-to-end using a physics-informed multi-objective loss function. This combines standard cross-entropy with stage-specific binary losses and an MIL consistency loss that penalizes disagreement between the globally assigned label, and the top-k attended spectra. Additionally, a peak-sensitive regularization (PSR) term is applied to spectral attention. By mathematically discouraging attention from concentrating on baseline fluctuations, the PSR guides the model to focus on chemically meaningful Raman shifts, improving the physical plausibility of its attributions. Collectively, this unified framework integrates spectral pattern recognition, weakly supervised instance selection, temporal dynamics, hierarchical decision-making, and peak-aware regularization for interpretable single-molecule peptide SERS analysis.

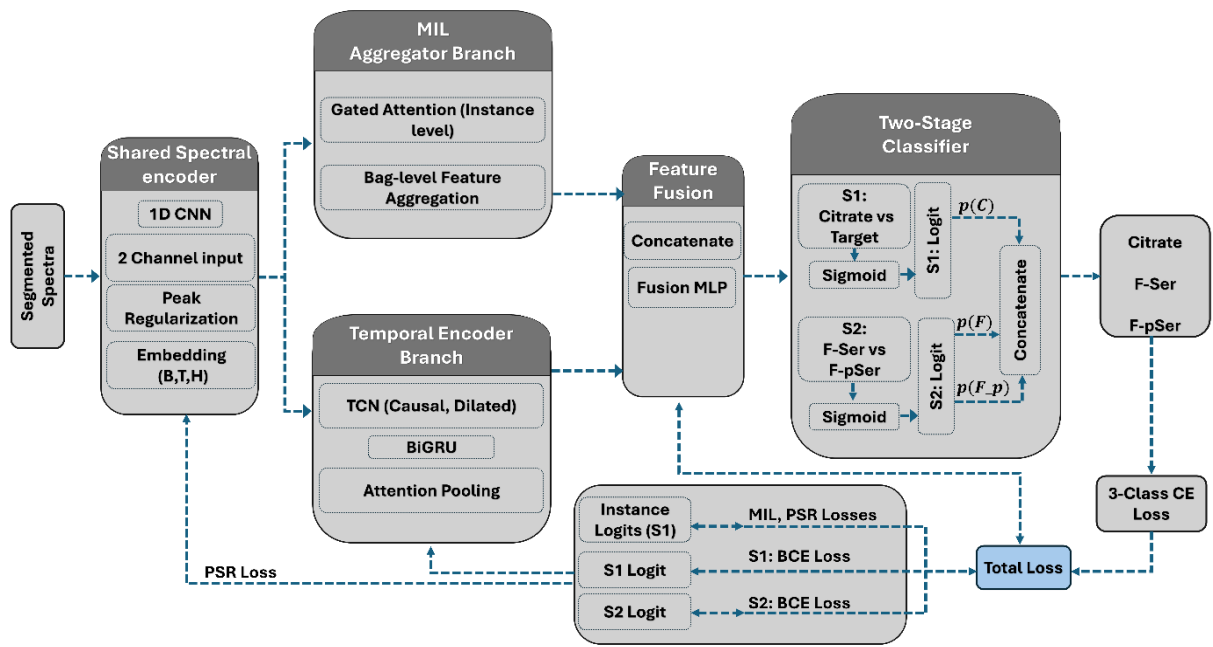

**Figure 4.** Bi-path hierarchical model for single-peptide PTM identification. SERS time-series segments are encoded by a shared Spectral Encoder, then split into two orthogonal branches: (i) a Multiple Instance Learning (MIL) aggregator with gated attention to extract sparse, high-intensity binding events and produce a bag-level feature, and (ii) a Temporal Encoder (causal dilated TCN, BiGRU, and attention pooling) to capture continuous binding dynamics. Fused features from the two branches are passed to a two-stage classifier that first separates citrate background from targets, then distinguishes F-Ser from F-pSer.

**Hierarcical classification Accuracy**

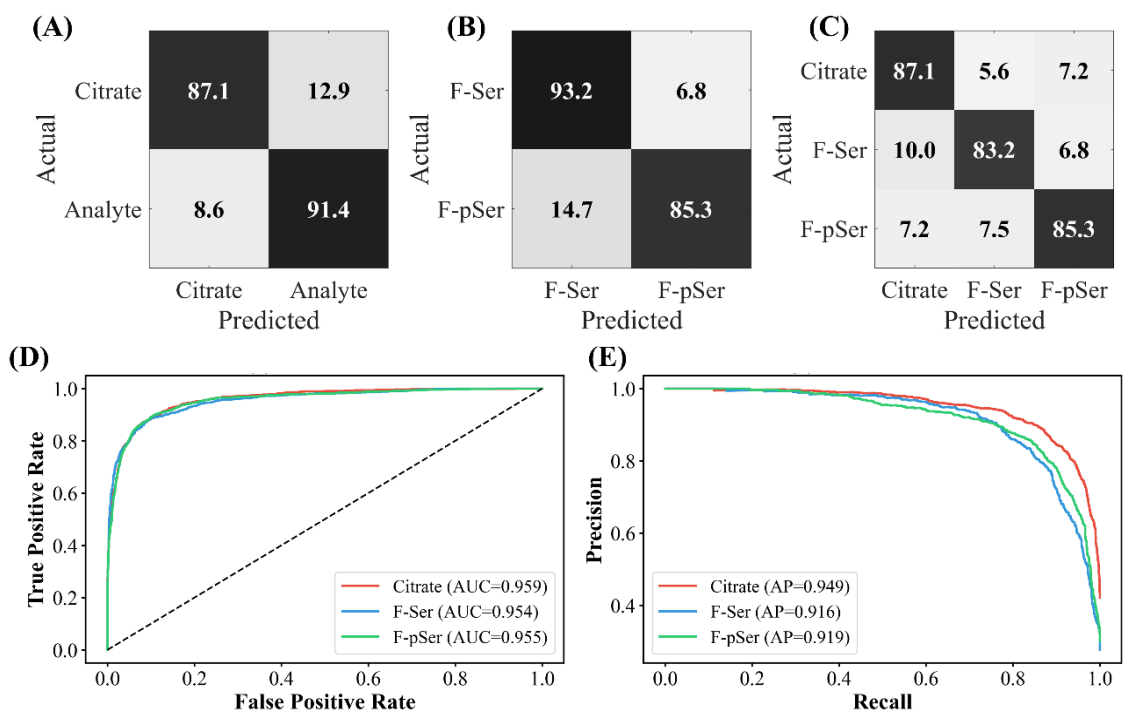

**Figure 5.** Performance evaluation of the Bi-Path SERS Classifier. Binary class confusion matrices for (A) the Stage 1 Citrate vs. Analyte classifier and (B) the Stage 2 F-Ser vs. F-pSer classifier. (C) The final aggregated three-class confusion matrix. The model's discriminative capability is further demonstrated by (D) the Receiver Operating Characteristic (ROC) curves and (E) the Precision-Recall (PR) curves, with their respective area under the curve (AUC and AUPRC) values.

To discriminate F-Ser from F-pSer, we employed a hierarchical classifier that first isolates true single-molecule analyte events from citrate background, and then resolves their subtle spectral differences. **Figure 5** summarizes the predictive performance of this Bi-Path architecture on an unseen dataset, utilizing hierarchical confusion matrices alongside Receiver Operating Characteristic (ROC) and Precision-Recall (PR) curves. The Stage 1 binary classifier effectively discriminates sparse molecular binding events from citrate interference, achieving true positive rates of 87.1% and 91.4% for citrate and analyte detection, respectively (**Figure 5A**). Conditioned on analyte presence, the Stage 2 classifier distinguishes F-Ser from F-pSer with accuracies of 93.2% and 85.3%, respectively (**Figure 5B**). When aggregated into a final three-class prediction (**Figure 5C**), the model maintains robust accuracies across all the classes (Citrate: 87.1%, F-Ser: 83.2%, F-pSer: 85.3%).

The model's performance is further supported by ROC curves (**Figure 5D**), which exhibit Area Under the Curve (AUC) values exceeding 0.95 for all classes. Because single-molecule SERS datasets are imbalanced by the rarity of true binding events, the Precision-Recall (PR) curves provide a more rigorous assessment of predictive reliability (**Figure 5E**). The model maintained a high Area Under the PR Curve (AUPRC) of above 91.6% for each classes, guaranteeing a remarkably low false discovery rate when identifying the specific post-translational modification. Together, these metrics demonstrate that the physics-informed hierarchical approach can successfully mitigates the profound signal ambiguity characteristic of single-molecule SERS.

**Physical Interpretability via 1D Integrated Gradients**

A critical limitation of deep learning models in spectroscopic and chemometric applications is the "black box" nature of neural networks, which often rely on statistically spurious dataset artifacts rather than genuine biochemical signatures. To verify our Bi-Path SERS Classifier relies on meaningful features, we evaluated the spatial attribution of the Spectral Encoder using 1D Integrated Gradients (1D-IG). Conceptually analogous to 1D Gradient-weighted Class Activation Mapping (1D-Grad-CAM) frequently used in recent analytical chemistry literature to decode 1D-CNNs[57,58], 1D-IG provides a high-resolution, feature-level attribution map by integrating the gradients of the output class probability with respect to the input Raman shift. By projecting these 1D-IG feature weights over the empirical peak occurrence frequency, we can directly observe whether the model's mathematical "attention" aligns genuine molecular vibrations.

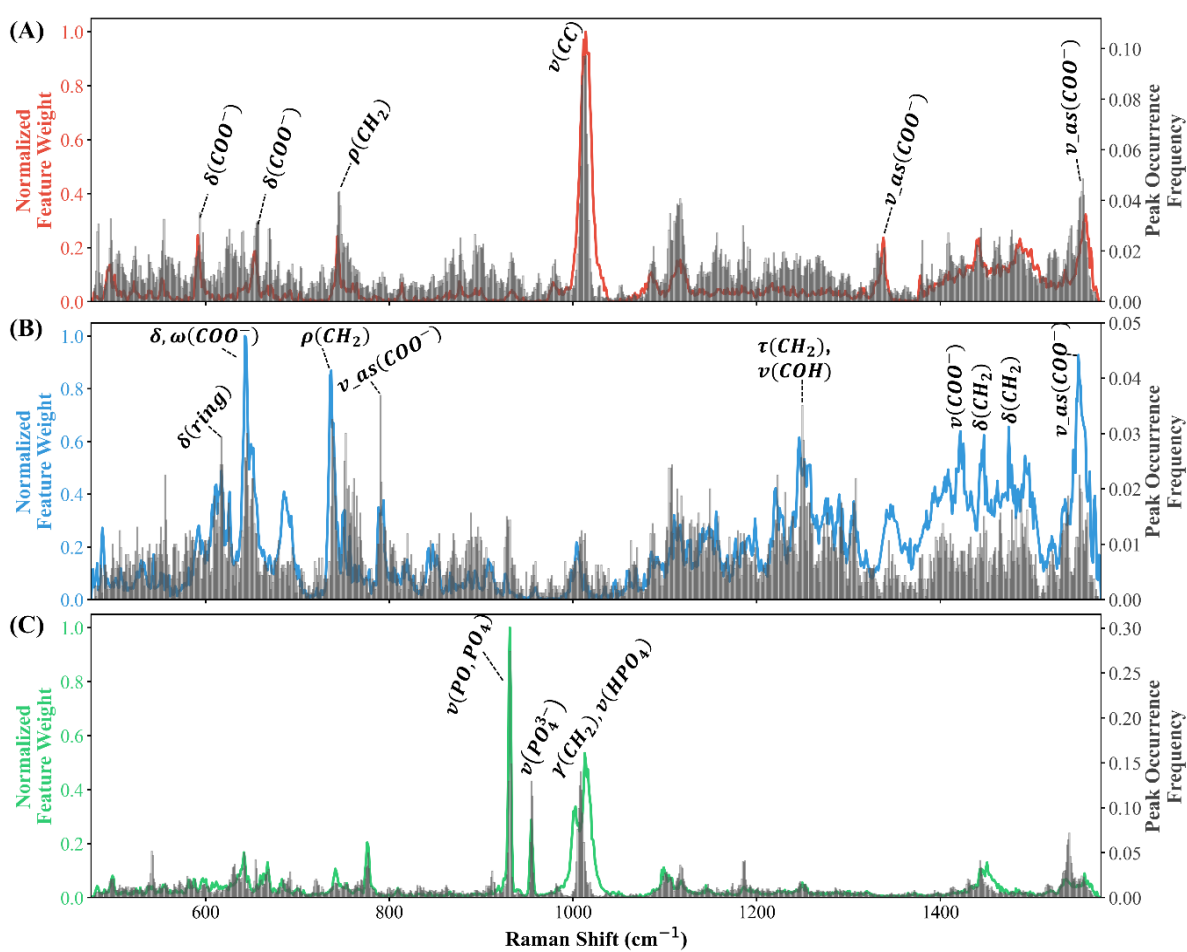

**Figure 6**. Integrated gradient curve and peak occurrence frequency for Citrate (a), F-Ser (b), and F-pSer (c) classes derived from the bi-path single-molecule SERS classification model. For each class, the solid-colored curve represents the 1D normalized feature attribution profile computed via Integrated

Gradients (IG), averaged across all test segments of each class. The bar graph overlaid on each panel represents the peak occurrence frequency of the fraction of class-specific spectra in which a Raman peak was detected at each wavenumber position. Spectral positions where high IG attributes co-localize with elevated peak occurrence frequency identify reproducible Raman vibrational bands, which are most discriminative for each molecular class.

The IG attribution curves of F-pSer and Citrate in **Figure 6** align with experimentally observed peak-occurrence frequency, which confirms the model's learned representations are grounded in genuine molecular signatures rather than spurious background artifacts. The IG profile for F-Ser also demonstrates strong agreement with its corresponding peak-occurrence frequency. Analysis of the feature map visualizations reveals that the model successfully identifies key vibrational modes. For example in Citrate, the model highlights the bending of the carboxylate group $\delta(COO^-)$ at 595 and 673 $cm^{-1}$, the carbon-carbon ($v(C-C)$) stretching vibration at 1015 $cm^{-1}$, and the asymmetric stretching of the carboxylate group ($v\_as(COO^-)$) at 1125, 1345, and 1550 $cm^{-1}$. In the case of F-Ser, we have observed bending, wagging and vibrations of carboxylate group at 642, 790, 1423 and 1550. Furthermore, it identifies ring bending modes at 620 and 1483 $cm^{-1}$, $CH_2$ rocking at 740 $cm^{-1}$, and torsion of $CH_2$ and vibration of C-OH at 1250 $cm^{-1}$. Similarly, for F-pSer, the model correctly isolates the vibrations characteristic of the phosphate group at 930, 950, and 1005 $cm^{-1}$, unequivocally demonstrating its ability to spatially localize the specific post-translational modification without human annotation. A comprehensive peak assignment for each class is provided in Table 2.

Table 2. Vibrational modes assigned to the SERS peaks shown in Figure 6, with references.

| Citrate | Vibrational mode | References |
|---|---|---|
| 595 | $\delta(COO^-)$ | 59 |
| 673 | $\delta(COO^-)$ | 59 |
| 1015 | $v(CC)$ | 59–62 |
| 1345 | $v\_as(COO^-)$ | 59,61 |
| 1550 | $v\_as(COO^-)$ | 59,60 |
| F-Ser | Vibrational mode | References |
| 620 | $\delta(ring)$ | 63 |
| 642 | $\delta, \omega(COO^-)$ | 64 |
| 740 | $\rho(CH_2)$ | 34 |
| 790 | $v\_as(COO^-)$ | 35 |
| 1250 | $\tau(CH_2), v(COH)$ | 34,35,37 |
| 1423 | $v(COO^-)$ | 65 |
| 1461 | $\delta(CH_2)$ | 65 |
| 1483 | $\delta(CH_2)$ | 63,64 |
| 1550 | $v\_as(COO^-)$ | 59,60 |
| F-pSer | Vibrational mode | References |
| 930 | $v(PO, PO_4)$ | 66–68 |
| 950 | $v(PO_4^{3-})$ | 69,70 |
| 1005 | $\gamma(CH_2), v(HPO_4)$ | 71,72 |

**Pathway Probability Score Analysis**

To quantitatively evaluate the discriminative capability and statistical certainty, we analyzed the probability score distributions across the three-classes (**Figure 7**) and their corresponding probability densities **(Figure S4** in the supporting information**)**. In each pathway, the model assigns high probabilities to the correct class while suppressing non-corresponding classes. Specifically, citrate samples clustered near to 1 in Figure 7A, with F-Ser and F-pSer near to 0. Similarly, F-Ser samples show high probability in Figure 7B, and F-pSer samples in Figure 7C, while the remaining classes are confined to low-probability regions.

From an interpretability perspective, these pathway-level probability distributions provide an intermediate explanatory layer between raw attributes and final classification outputs. While integrated gradients in **Figure 6** highlight the specific Raman shift regions contributing most to a prediction, the pathway probability scores reflect how confidently the model supports each biochemical hypothesis.

In highly sensitive and stochastic analytical techniques such as single-molecule peptide SERS, deterministic, threshold-based predictions are insufficient. Reliable diagnostics require well-calibrated probability estimates to prevent transient noise from being misinterpreted as genuine molecular trapping events. This hierarchical framework mirrors the decision-making process of human experts and mitigates overconfident misclassifications, a recognized bottleneck in SERS diagnostics. Recent studies employing MIL for SERS-based liquid biopsies has demonstrated that sequential hierarchical architectures where baseline anomaly detection strictly gates subsequent fine-grained sub-classifications are essential for achieving high-fidelity molecular identification[50].

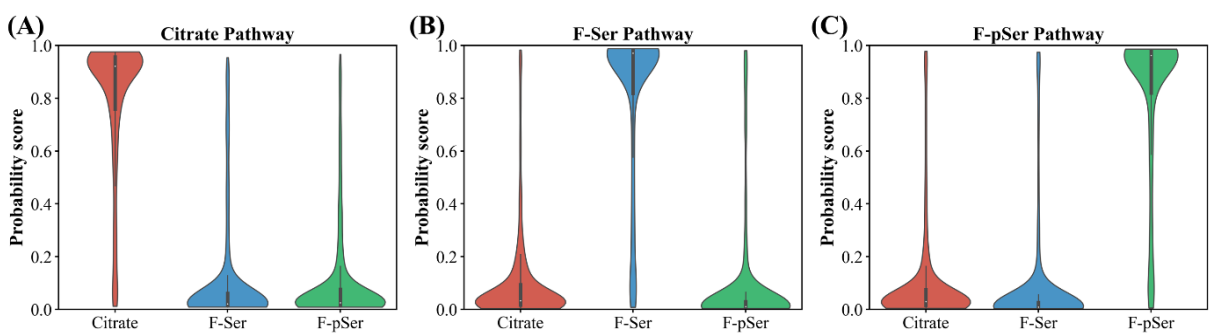

**Figure 7.** Pathway Confidence Analysis of the Bi-Path hierarchical deep learning model. The violin plots of probability scores of Citrate pathway (A), F-Ser pathway (B), and F-pSer pathway (C). The figure illustrates the distribution of final probability scores generated by the hierarchical classification head across the test dataset.

## Conclusion

In conclusion, we establish a physics-informed, bi-path hierarchical deep learning framework to decode highly stochastic single-molecule SERS (SM-SERS) signals from a particle-in-pore plasmonic sensor. This approach addresses key limitations of SM-SERS by extracting chemically meaningful signatures from temporally fluctuating, low-intensity, and background-contaminated spectra, where conventional analytical and supervised learning methods often fail.

The framework integrates correlation-guided segmentation, attention-based multiple instance learning (MIL), and a temporal encoder to jointly capture spatial and temporal features of dynamics peptide signals. In SM-SERS peptide measurements, spectral sequences contain transient binding events, idle states, citrate interference, and partial hotspot excitation, making frame-level annotation impractical. This challenge is overcome through adaptive segmentation and MIL, enabling the model to identify and emphasize rare, information-rich spectra. In parallel, the temporal encoder (TCN and BiGRU) captures blinking dynamics and sequential dependencies associated with analyte diffusion through the plasmonic hotspot.

The fusion of spatial and temporal features, followed by hierarchical classification, enables effective suppression of dominant background signals and accurate discrimination of peptides with subtle spectral differences. Peak-sensitive regularization further constrains the model to focus on Raman-active features, reducing sensitivity to noise and baseline artifacts, while 1D-IG attribution enhances chemical interpretability. This unified approach enables detection of subtle post-translational modification signatures within highly overlapping spectral backgrounds.

Overall, this work establishes a robust computational paradigm for single-molecule chemometrics, demonstrating that weakly supervised, physics-aware deep learning can transform stochastic plasmonic sensing into an interpretable molecular readout. By filtering complex backgrounds and resolving transient sub-molecular variations, the framework holds strong translational potential for next-generation liquid biopsies and real-time biological profiling. Beyond peptide phosphorylation, this architecture will seamlessly translate into broader clinical diagnostics, including early-stage cancer and neurodegenerative biomarker analysis. Ultimately, our core mechanism for decoding label-ambiguous temporal data can be adapted across single-molecule biophysics and medical imaging, paving the way for ultra-sensitive, real-time diagnostic tools in precision medicine.

## Methods

### Material

We purchased F-Ser and F-pSer from Sigma-Aldrich. MicroChemicals GmbH supplied silicon wafers with 100 nm $Si_3N_4$ membranes. The nonfunctionalized 50 nm gold nanoparticles (AuNPs) from Sigma (concentration of $3.5 \times 10^{10}$ particles/mL) in 0.1 mM PBS citrate stabilized, reactant free.

### Apparatuses and Software used

We acquired SERS using a Thermo Fisher DXR2xi Raman imaging microscope equipped with an Andor Solis software for over 200 s, over which a 0.1 s exposure time for each spectrum. The nanopore was fabricated via focused ion beam (FIB) milling and characterized by a Sigma HD VP field-emission scanning electron microscope (FE-SEM). A gold layer was deposited using a Q150T ES sputter coater. Spectral preprocessing was performed in MATLAB R2022b (See supporting information **Figure S1**), while subsequent data organization and model training and validation conducted by using Python within the Visual Studio Code environment. To optimize computational efficiency for the deep learning pipeline, model training was accelerated using NVIDIA CUDA on a GPU.

**Nanopore fabrication**

The gold nanopores were fabricated by using Focused Ion Beam (FIB) on commercial $Si_3N_4$ membranes supported on silicon. The size and thickness of the SiN window was 1 × 1mm and 100nm respectively. After sputtering a 2 nm titanium and 100 nm gold layer on the front side and 2nm gold layer at back side of the $Si_3N_4$ membrane, FIB milling (FEI Helios DualBeam) from the back side of the membrane to create nanopores with 200nm diameter.

**Spectra Preprocessing**

Cosmic ray artifacts were first removed using median filtering in MATLAB. Baseline correction was subsequently performed using the asymmetric least squares (ALS) method to suppress fluorescence and background effects. The spectra were then denoised using a Savitzky–Golay filter (window = 9, polynomial = 3) and standardized via min–max normalization. Finally, a Pearson correlation algorithm was used to isolate particle-trapping events from empty nanopore noise, and the data were segmented for bi-path hierarchical modeling.

**SERS Spectra Segmentation**

To resolve stochastic spectral fluctuations from peptide diffusion and blinking, we employed Pearson correlation-based segmentation. The algorithm iteratively compares time-series spectra to a reference (**$S_{ref}$**), grouping similar signals into "bags" to capture discrete molecular states. Once a valid $S_{ref}$ is established at index $i$, the algorithm initiates a forward-looking temporal window (24 frames) which is consistent with a ~2.5 s resident-time of the molecule in the hotspot. For each subsequent frame **$S_k$** (where $k > i$ ), the Pearson Correlation coefficient (r) calculated as follows:

$$r = \frac{\sum(S_{ref}-\bar{S}_{ref})(S_k-\bar{S}_k)}{\sqrt{\sum(S_{ref}-\bar{S}_{ref})^2 \sum(S_k-\bar{S}_k)^2}} \quad (1)$$

We considered a correlation threshold $r > .65$ and a segment containing less than 3 spectra discarded as stochastic noise. This approach isolates high-fidelity molecular signatures from noise and idle states by prioritizing spectral morphology over absolute intensity.

**Training of the model**

The deep learning model was trained using a supervised hierarchical approach designed to resolve single-molecule SERS signatures across three distinct classes: Citrate, F-Ser, and F-pSer. To address the class imbalance, we implemented a Weighted Random Sampler and applied computed class weights to a specialized Two Stage Loss function (detailed in the GitHub link), which jointly optimized classification accuracy and the hierarchical decision-making process. The training utilized the AdamW optimizer with an initial learning rate of $3*10^{-4}$ and a weight decay of $10^{-4}$ to prevent overfitting. We employed a ReduceLROnPlateau scheduler to adaptively adjust the learning rate based on validation performance, complemented by a five-epoch linear warmup phase to stabilize early-stage gradient flow. To manage the computational load of high-dimensional spectral segments (up to 25-time steps), we utilized gradient accumulation with a total effective batch size of 64 and implemented mixed-precision training on an NVIDIA GPU. The model was trained for 100 epochs, with an early stopping patience of 20 epochs based on validation accuracy to ensure the selection of the most robust and generalized parameter state.